# Leakage-Resistant Evaluation of AI-Generated Database Specifications in Specification-Driven Migration

Oleg Grynets
EPAM Systems
McLean, Virginia, USA
oleg_grynets@epam.com

Ivan Fedorchuk
EPAM Systems
Vinnytsia, Ukraine
ivan_fedorchuk@epam.com

Vasyl Lyashkevych
EPAM Systems
Lviv, Ukraine
vasyl_lyashkevych@epam.com

***Abstract*—This paper presents a leakage-resistant framework and benchmark (LRFB) for measuring evaluators of AI-generated Oracle PL/SQL specifications. The benchmark contains 494 labelled source–specification pairs across 19 quality metrics grouped into seven dimensions; each metric contains 26 records divided into development and held-out subsets of 13 records each, with three clean and ten defective specifications per subset. Ground truth is mechanically sealed from inference and checked with canary and label-permutation tests. Three LLM judges were evaluated with three runs per record and compared against two trivial baselines. On 247 held-out records, GPT-4o, DeepSeek V4 Pro, and Claude Sonnet 4.5 reached balanced accuracies of 0.87, 0.89, and 0.81, respectively, while a specification-length baseline reached 0.88. Exact paired tests did not separate any judge from that baseline. The judges nevertheless provided information unavailable to length: the labelled defect category appeared among findings in 84%–92% of violated records, and quoted evidence overlapped the labelled span in 92%–93%. The study therefore reframes specification evaluation from binary issue detection toward evidence-linked, typed findings and benchmark-validity controls.**

## I. Introduction

Large language models (LLMs) increasingly participate in software modernization, not only by translating code but also by creating intermediate representations that make transformation steps explicit and reviewable. Specification-based Code–Text–Code reengineering treats a technology-neutral specification as a controllable boundary between source and regenerated code [1]. In Oracle-to-PostgreSQL migration, this view has been extended to token-aware migration strategies [2] and to cross-agent specification portability, where the same intermediate representation may be consumed differently by distinct development agents [3]. Specification-driven DevOps similarly shows that executable artifacts can be generated while still omitting deployment intent that is not explicit in the repository [4].

This trajectory changes the assurance target. Once a specification becomes an operational artifact, its quality is no longer merely documentation quality: it controls what downstream agents regenerate, validate, deploy, and monitor. Security-oriented specification-driven work has shown that implicit requirements can disappear during generation unless they are represented and verified explicitly [5]. Risk studies of LLM-based software development likewise identify semantic drift, weak traceability, and unreliable transformations as lifecycle-level concerns [6], while taxonomy-based risk modelling connects requirements and development activities to structured risk knowledge [7].

Database migration is an especially suitable setting for studying this problem. A recovered specification may look fluent while omitting an OUT parameter, a transaction side effect, a NULL-sensitive condition, a trigger timing constraint, or a source-dialect semantic that must be surfaced before migration. Conversely, the specification may invent an operation that the source never performs. If such a specification is used as the generation contract for PostgreSQL, the resulting target code may be syntactically valid but behaviorally wrong.

The central problem is therefore not only how to generate a specification, but how to measure an evaluator that claims to determine whether a generated specification is faithful, complete, neutral, internally consistent, clear, and migration-aware. Existing reference-free evaluation methods provide useful building blocks, existing approaches provide useful foundations for specification evaluation, including claim-based factual consistency [8], [9], [11], question-answering verification [12], LLM-based judging [13]–[17], and code–text alignment metrics [18], [19]. However, these approaches primarily assess generated artifacts rather than systematically testing evaluators against labelled specification defects under leakage-resistant experimental conditions. However, these methods usually estimate a property of the generated text; fewer studies ask whether the evaluator itself can locate a known defect under controlled conditions.

This paper addresses that measurement problem with LRFB. It does not introduce a new specification generator and does not claim that an LLM judge is intrinsically superior to deterministic or embedding-based methods. Instead, it introduces a benchmark and experiment framework designed to test evaluators without leaking labels, to compare them with trivial shortcuts, and to measure more than a binary verdict. The contribution is deliberately methodological because two threats materialized during the study: the original record identifiers encoded the labels, and specification length alone predicted many benchmark labels.

The paper makes four contributions. First, it defines a 19-metric quality system for AI-generated Oracle PL/SQL specifications. Second, it provides a 494-pair silver benchmark with development and held-out halves and a strict ground-truth boundary. Third, it reports a controlled comparison of three LLM judges against always-issue and specification-length baselines. Fourth, it shows why category, affected-item, and evidence agreement are more informative than issue detection when a surface heuristic can match the judges.

As illustrated in Fig. 1, the proposed workflow introduces a specification quality evaluation stage between the recovery of a technology-neutral specification from legacy Oracle artifacts and its subsequent implementation in the target database dialect. The evaluation stage identifies evidence-linked defects using source–specification pairs, while target-code generation remains outside the scope of this study.

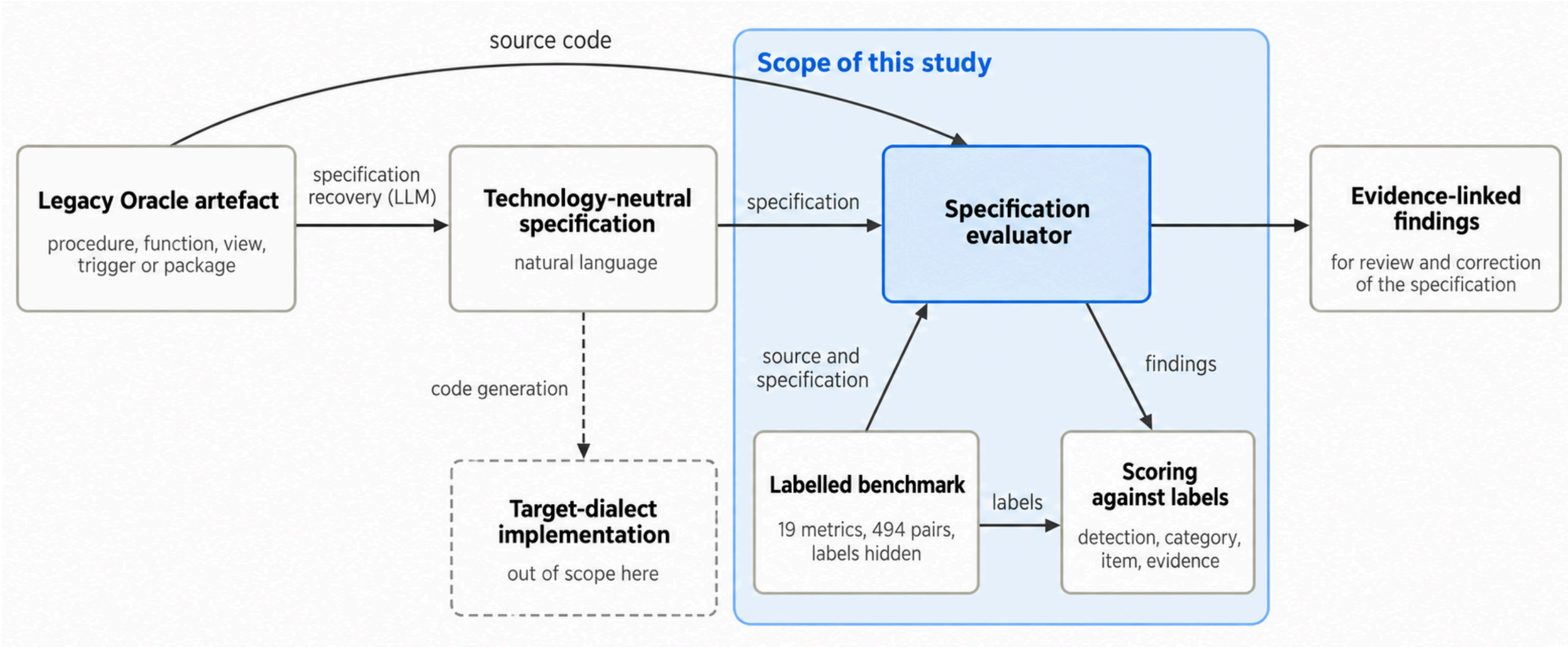

Fig. 1. Specification-driven database migration and the scope of this study. The specification is recovered from a legacy Oracle artifact and evaluated before target-dialect generation; target-code generation is outside the reported experiment.

The workflow establishes a clear boundary between specification recovery, quality evaluation, and target-code generation. By comparing the recovered specification with the original source code, the evaluator produces evidence-linked findings that support defect identification and subsequent correction. The labelled benchmark enables systematic assessment of detection accuracy, defect categorization, affected-item identification, and evidence agreement without exposing ground-truth labels during inference.

## II. Research Trajectory and Related Work

### A. From Specification-Mediated Migration to Specification Assurance

The present study is positioned as the evaluation layer in a broader specification-driven software-engineering trajectory. Specification-based Code–Text–Code reengineering [1] motivates the intermediate specification as an explicit representation of behavior, identifiers, control flow, side effects, data dependencies, and domain intent. Token optimization work [2] shows that migration quality depends on what semantic information is preserved when the representation is compressed. Cross-agent portability experiments [3] further show that specification artifacts cannot automatically be assumed to be agent-neutral: the same specification can lead to agent-dependent degradation. These results motivate treating the specification as a first-class object that requires its own quality assurance before reuse.

The same principle appears outside database code. Specification-driven DevOps [4] distinguishes functional executability from deployment-intent fidelity; generated Docker artifacts can run while omitting architectural or production constraints. Security knowledge transition work [5] operationalizes otherwise implicit security requirements as explicit, traceable elements in a specification. Complementary studies of LLM-related software risks [6] and lifecycle-oriented risk taxonomies [7] argue that AI-mediated software development requires explicit risk representation, validation, and traceability. Taken together, these works move the research question from “can an LLM generate an artifact?” to “what information must be represented, preserved, checked, and evidenced before that artifact can safely drive the next lifecycle stage?”

### B. Reference-Free Faithfulness and Completeness

Reference-free factuality research provides a natural starting point for specification assurance. RAGAs decomposes generated answers into statements and estimates whether each is supported by retrieved context [8]. RAGEval adds scenario-specific datasets and separate completeness, hallucination, and irrelevance measures [9].

A software-engineering RAG evaluation methodology emphasizes strong baselines, task-specific metrics, controlled experimental design, and explicit reporting of design decisions [10]. FActScore similarly decomposes a generation into atomic facts and estimates the proportion supported by a trusted source [11]. These methods correspond closely to specification faithfulness, but they are precision-oriented: a very short specification can be perfectly faithful while omitting most of the source behavior.

Question-answering methods address part of the omission problem. QAFactEval generates questions from one text, answers them from another, and compares the answers [12]. Reversing the direction—from source to specification—makes the method recall-like and therefore suitable for coverage. The difficulty is that question generation adds another model-mediated step, and the number of questions needed to cover a database artifact is itself an evaluation problem.

### C. LLMs as Judges

G-Eval demonstrated that criterion-conditioned LLM evaluation can correlate well with human judgment on natural-language generation [13]. Subsequent surveys document the rapid expansion of LLM-as-a-judge methods together with persistent concerns about prompt sensitivity, position bias, verbosity bias, self-preference, calibration, and run-to-run variability [14], [15]. In software engineering, CodeRPE asks LLMs to evaluate code summaries from multiple roles and reports strong correlation with human ratings [16]. ReFEree goes further by producing typed and localised factual-consistency findings for code summaries using dependency context [17]. These works motivate the structured findings used in LRFB, but they also make evaluator validation essential: an explanation is useful only if the verdict, category, location, and evidence are themselves reliable.

### D. Code–Text Alignment and Program Representations

Embedding-based metrics provide deterministic alternatives. CodeBERTScore adapts contextual token matching to code generation [18], while SIDE learns a contrastive code-summary representation whose similarity captures semantic alignment independently of a reference summary [19].

The representation literature also shows why purely lexical comparison is insufficient. CodeBERT learns joint natural-language and programming-language representations [20]; GraphCodeBERT incorporates data flow [21]; UniXcoder combines code, comments, and abstract syntax structure [22]; and PLBART supports program understanding and generation through sequence-to-sequence pre-training [23]. CodeBLEU combines lexical overlap with syntax and data-flow components [24]. These methods are valuable for global alignment but normally return a scalar; migration review instead requires a specific finding such as “the ELSE branch is missing” or “the specification invents a COMMIT.”

### E. Evaluation in Code Generation and Text-to-SQL

Program-generation benchmarks reinforce the importance of executable or semantic oracles. HumanEval evaluates generated Python with unit tests [25], AlphaCode uses competitive-programming problems to measure synthesis at higher difficulty [26], and SWE-bench grounds evaluation in real repository issues and execution environments [27]. Database-oriented work has analogous concerns: Spider established cross-domain text-to-SQL evaluation on unseen schemas [28], while PICARD constrains language-model decoding with incremental parsing to reduce invalid SQL [29].

LRFB differs in target: it evaluates a natural-language representation recovered from existing source code rather than the target code itself. Nevertheless, these benchmarks support the same methodological principle—evaluation should be tied to a task-specific oracle rather than surface plausibility.

### F. Long Context, Shortcut Signals, and Benchmark Validity

Long-context behavior and prompt compression are directly relevant because real legacy artifacts and their dependency context can exceed convenient prompt sizes. “Lost in the Middle” demonstrates that models may use long contexts unevenly depending on where relevant information appears [30]. LLMLingua [31] and LongLLMLingua [32] show that prompt compression can reduce cost, but preservation of task-relevant information is a design constraint rather than a guaranteed side effect. SelfCheckGPT detects hallucination through cross-sample inconsistency without external knowledge [33], while MT-Bench and Chatbot Arena show both the scalability and characteristic biases of LLM judges [34].

More generally, evaluation research has repeatedly shown that a metric can correlate with desired quality for the wrong reason. BERTScore provides a semantic alternative to lexical overlap [35]; SummaC [36], FactCC [37], and FEQA [38] expose the difficulty of factual-consistency measurement; and surveys of hallucination in natural-language generation emphasize the need to distinguish faithfulness, factuality, and task-specific grounding [39]. LRFB contributes an analogous approach for specification benchmarks: a high detection score is not interpretable until label leakage and trivial correlates have been ruled out.

### G. Research Gap

The reviewed literature leaves four connected gaps. First, specification-mediated migration needs explicit evaluation of the intermediate representation, not only the regenerated code. Second, most reference-free metrics evaluate text quality but do not test the evaluator on known defects with typed labels. Third, omission errors require recall-oriented measures alongside faithfulness. Fourth, benchmark validity must be tested against label leakage and shortcut baselines. LRFB addresses these gaps by measuring the evaluator rather than proposing another generation metric.

## III. PROBLEM FORMULATION

### A. Source Facts and Specification Claims

Let c denote a source artifact and s a natural-language specification recovered from it. Let $F(c)$ be the set of relevant source facts and $A(s)$ the set of atomic claims made by the specification. Writing $s \vDash f$ when the specification expresses source fact f and $c \vDash a$ when the source supports specification claim $a$, two complementary quantities are:

$$Coverage(c,s) = \frac{|\{f \in F(c) : s \vDash f\}|}{|F(c)|}. \quad (1)$$

$$Faithfulness(c,s) = \frac{|\{a \in F(s) : c \vDash a\}|}{|F(s)|}. \quad (2)$$

Here, $c$ is the source artifact, $s$ is its specification, $F(c)$ is the set of relevant source facts, and $A(s)$ is the set of atomic specification claims.

Coverage decreases when source behavior is omitted; faithfulness decreases when the specification asserts unsupported behavior. Neither quantity alone is sufficient. A specification that describes only the procedure name may be faithful but uselessly incomplete; a verbose specification may have high coverage while introducing unsupported claims.

Table I summarizes the main approaches to evaluating generated specifications against their source artifacts, highlighting differences in evaluation objectives, verification mechanisms, output granularity, and the ability to detect omitted information.

The comparison reveals that existing approaches primarily focus on factual consistency, semantic similarity, or holistic quality assessment, while providing limited support for detecting omissions and producing localized, evidence-linked findings. These limitations motivate the proposed evaluation framework, which combines 19 quality metrics with defect-level assessment of detection accuracy, category identification, affected items, and supporting evidence.

## IV. Method and Experimental Design

### A. Metric System

The benchmark evaluates seven specification-quality dimensions, refined into 19 metrics. Each metric has a closed vocabulary of ten issue categories.

The evaluator receives the source code, the generated specification, the metric definition, and the category vocabulary. This design makes category-level scoring possible and avoids post-hoc interpretation of free-form error labels.

Table II provides the complete classification of the evaluation metrics, while Fig. 2 illustrates their conceptual organization into seven quality-assessment components. These groups address source coverage, faithfulness, completeness, specification neutrality, internal consistency, clarity, and migration-related risks.

TABLE I. Position of LRFB relative to representative evaluation approaches

| Approach | Primary target | Decision mechanism | Output | Omissions |
|---|---|---|---|---|
| RAGAs [8] | Faithfulness | LLM per statement | Ratio | No |
| RAGEval [9] | Coverage + hallucination | LLM per key point | Ratios | Yes |
| FActScore [11] | Factual precision | Atomic-fact support | Ratio | No |
| QAFactEval [12] | Faithfulness / coverage | Question answering | Score | If reversed |
| G-Eval [13] | Criterion quality | Holistic LLM judge | Rating | Criterion dependent |
| CodeRPE [16] | Code-summary quality | Role-based LLM judge | Ratings | Indirectly |
| CodeBERTScore [18] | Code-text alignment | Contextual embeddings | P/R/F | Aggregate only |
| SIDE [19] | Code-summary alignment | Contrastive similarity | Scalar | No |
| ReFEree [17] | Factual consistency | Segment-level LLM | Typed findings | Primarily precision |
| LRFB | Evaluator accuracy on 19 metrics | Labelled source–spec pairs | Detection + typed evidence | Yes |

TABLE II. Metric groups and metrics

| Group | Metrics |
|---|---|
| Source coverage | Source Fact Coverage; Critical Source Fact Coverage; Source Question Answerability |
| Faithfulness | Unsupported Claims; Contradicted Claims; Hallucinated Entities; Evidence Attachment |
| Completeness | Interface; Data Transformation; Control-Flow; Side-Effect Completeness |
| Neutrality | Source Dialect Leakage |
| Internal consistency | Entity Consistency; Behavioral Contradictions; Terminology Drift |
| Clarity | Ambiguous Statements; Unresolved References |
| Migration risk | Semantic Migration Risks; Risk Explanation Completeness |

The classification provides a multidimensional basis for identifying defects in AI-generated specifications. In particular, source coverage and faithfulness distinguish omitted source facts from unsupported claims, while completeness and consistency metrics capture missing behavioral details and conflicting statements. The remaining groups address technology neutrality, interpretability, and migration risks, enabling systematic evaluation of specification quality before target-code generation.

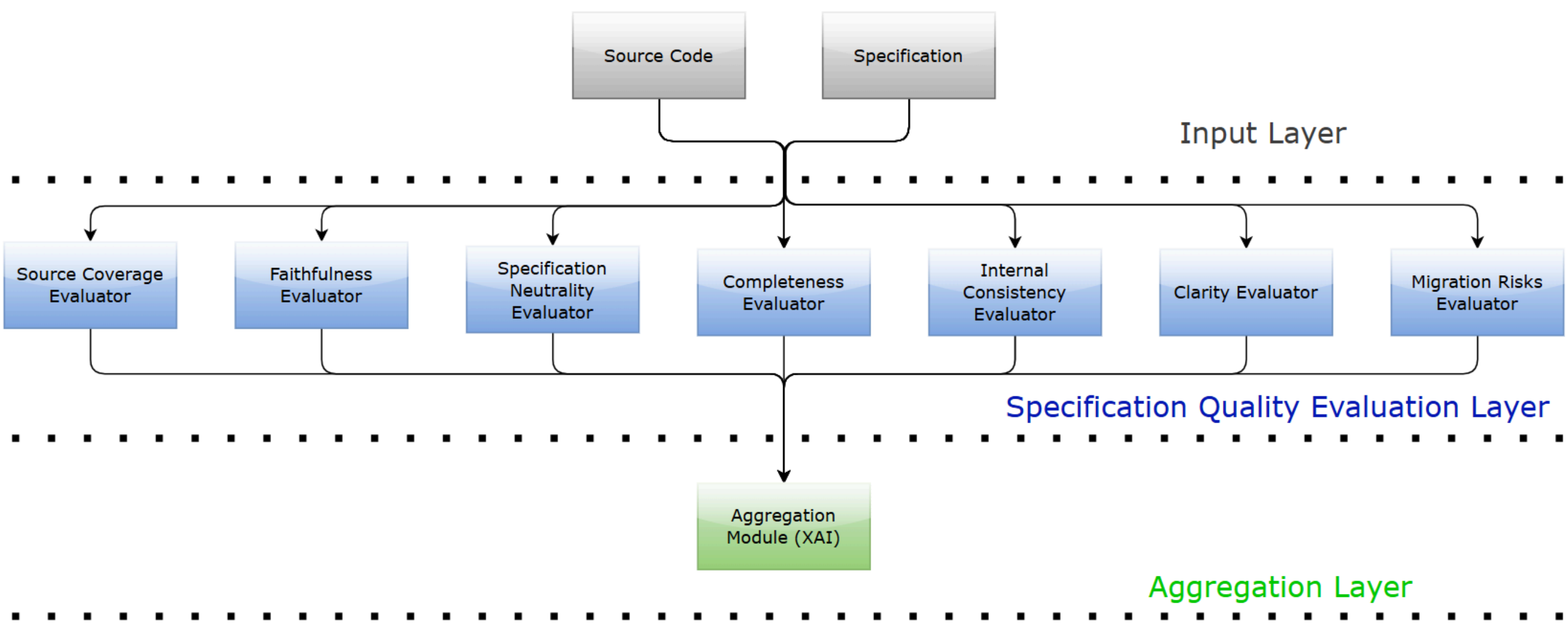

Fig. 2. Conceptual specification-quality evaluation layer. Source code and specification feed seven evaluator groups whose findings can later be aggregated. The reported experiment evaluates the quality-assurance layer; downstream aggregation policy is not evaluated.

The conceptual architecture in Fig. 2 groups the 19 specification-quality metrics into seven evaluation dimensions. Their outputs may subsequently be combined by an aggregation module; however, the present experiments evaluate individual metric-specific judgments and do not assess aggregation strategies or explainability mechanisms.

## B. Dataset Construction

Each of the 19 metrics has a synthetic dataset of 26 source–specification pairs, for 494 records in total. A source artifact is an Oracle procedure, function, view, trigger, or package of 4–29 lines (median 292 characters). A specification contains one to three sentences (median 133 characters). Six records per metric are labelled clean. Twenty contain exactly one deliberately introduced defect: two instances for each of ten issue categories. Labels contain the issue category, description, difficulty, object type and, when applicable, affected item, evidence span, and severity.

Each metric dataset is divided into development and held-out halves of 13 records, with one record per issue category and three clean records in each half. Because 28 source snippets are reused, any records sharing a source are assigned to the same half to reduce direct leakage across the split. The benchmark is therefore a controlled silver dataset, not a claim about the natural prevalence of defects in industrial specifications.

## C. Ground-Truth Boundary

Only inference inputs are allowed to reach a judge: source code, specification, metric name, and the metric-specific question for Source Question Answerability. All other fields are ground truth and remain in the scoring stage. This boundary became necessary because the original record identifiers encoded their labels: clean records contained a clean marker and violated records contained the exact issue category. LRFB therefore replaces identifiers with a hash of inference inputs and sends no record identifier to the judge. Leakage is tested rather than assumed. A canary dataset inserts a unique sentinel into every ground-truth cell; the test suite asserts that no sentinel appears in requests, inference logs, or model-output tables. A label-permutation test rewrites all labels and verifies that model requests and inference outputs remain byte-identical. These checks make the boundary an executable property of the framework.

The experimental pipeline separates inference inputs from ground-truth annotations to prevent label leakage during specification evaluation. As illustrated in Fig. 3, only source code, specifications, metric definitions, and metric-specific questions are accessible to the LLM judge, while ground-truth labels are reserved exclusively for the scoring stage.

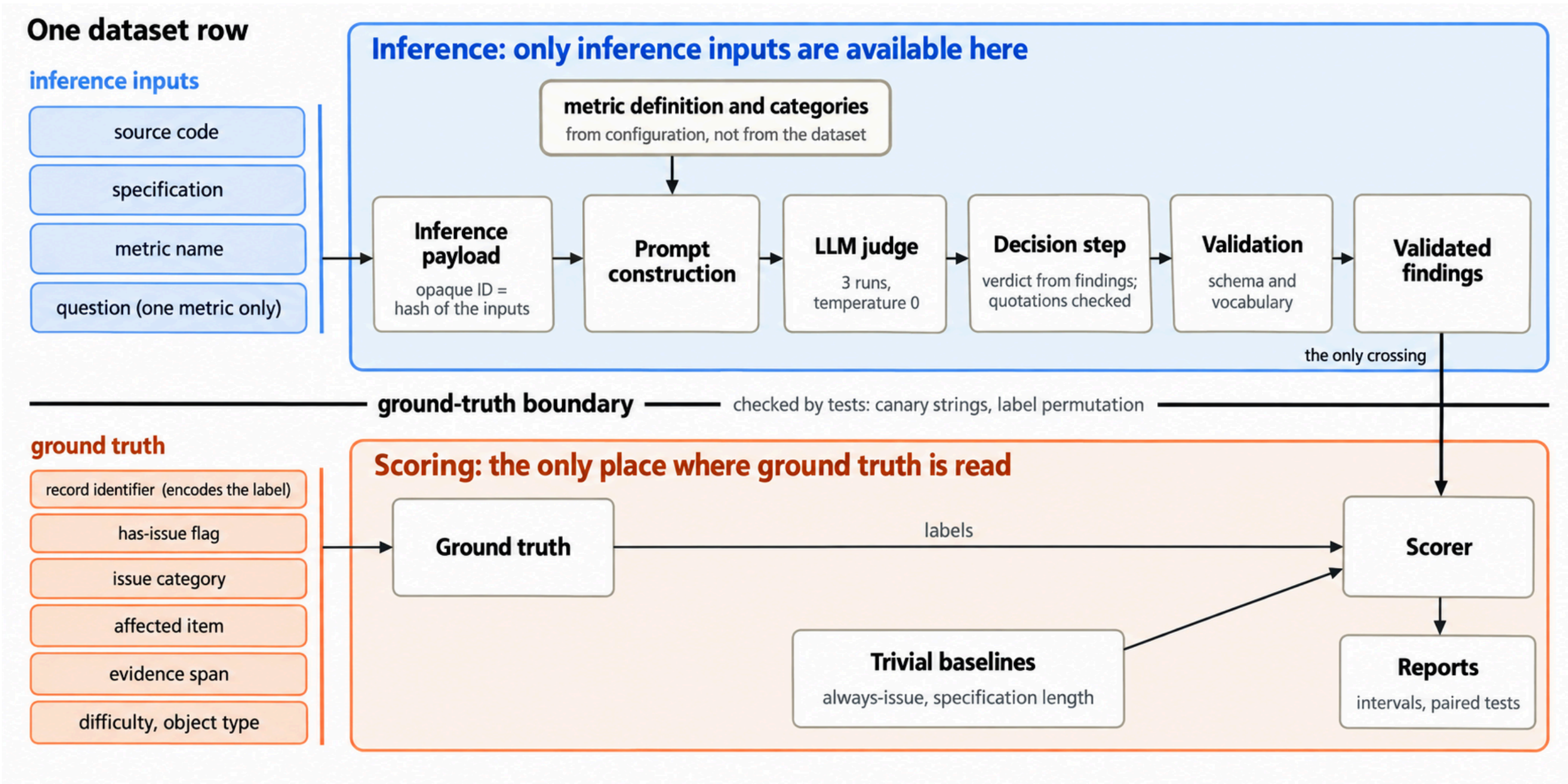

Fig. 3. Experiment pipeline and ground-truth boundary. Inference uses only source/specification inputs and metric configuration. Labels enter only at scoring. Canary and label-permutation tests verify that ground truth does not cross the boundary.

The resulting pipeline makes label isolation an explicit and testable property of the evaluation framework. Validated predictions are subsequently compared with ground-truth annotations only during scoring, preserving the independence of inference from benchmark labels.

## D. LLM Judge and Structured Findings

The judge receives the metric definition, ten category names with one-sentence definitions, source code, specification, and a JSON-compatible output schema. Each finding contains a category, affected item, severity, source/specification quotation where applicable, and a one-sentence description. The same prompt structure is used for all metrics; only the metric definition and categories change.

A deterministic decision layer converts the structured result into the binary issue verdict used for detection: a record has an issue if and only if the validated finding list is non-empty. Every quotation is checked against the exact input text and classified as grounded, missing, or ungrounded. Responses outside the schema or category vocabulary are invalid rather than silently repaired. The framework also contains deterministic and fact-grounded hybrid evaluator prototypes for interface facts, but they are not included in the reported comparison.

## E. Measures

Each configuration is evaluated three times per record. The final verdict is determined by majority voting among valid responses. In cases where no strict majority exists, the record is treated as unresolved and scored as incorrect. Records with no valid responses are also counted as incorrect. This rule ensures deterministic aggregation of repeated judgments. With *TP*, *FP*, *FN*, and *TN* defined over records, the primary detection measures are sensitivity, specificity, balanced accuracy, and Matthews correlation coefficient (MCC):

$$Sens = \frac{TP}{TP+FN},$$

$$Spec = \frac{TN}{TN+FP}, \quad (3)$$

$$BA = \frac{Sens+Spec}{2}.$$

$$x1 = TP + FP, \qquad x2 = TP + FN,$$

$$x3 = TN + FP, \qquad x4 = TN + FN,$$

$$MCC = \frac{TP \cdot TN - FP \cdot FN}{\sqrt{x1 \cdot x2 \cdot x3 \cdot x4}}, \quad (4)$$

Here, *TP*, *TN*, *FP*, and *FN* denote true positives, true negatives, false positives, and false negatives, respectively.

Balanced accuracy is preferred to ordinary accuracy because 20 of 26 records in each metric dataset are violated. An evaluator that always reports an issue therefore achieves 0.77 raw accuracy but only 0.50 balanced accuracy.

Beyond detection, LRFB scores whether the labelled category appears anywhere in the findings, whether it is the first finding, whether the affected item is identified by at least 0.5 content-token recall, and whether the quoted evidence overlaps the labelled span after case and whitespace normalization. These measures correspond to what a human reviewer actually needs: a typed, localised, evidenced claim about the specification. Proportions are reported with 95% Wilson intervals [40]. Composite measures use percentile bootstrap intervals with class-stratified record resampling. Paired correctness between two evaluators is compared with the exact McNemar test [41].

## F. Trivial Baselines

Two deliberately weak baselines are evaluated on the same records. Always-issue labels every record violated. Specification length labels a record violated when the specification character count is below a per-metric threshold fitted on the development half and then frozen for held-out evaluation:

$$\hat{y}_{len} = 1[|s| < 0] \quad (5)$$

The length baseline is not intended as a serious evaluator. It is a benchmark-validity probe: if it performs like an LLM that reads both source and specification, detection accuracy alone cannot demonstrate source-grounded reasoning.

## G. Experimental Protocol

Three model configurations were evaluated: GPT-4o (2024-11-20 snapshot), Claude Sonnet 4.5 (2025-09-29), and DeepSeek V4 Pro (API identifier: deepseek-v4-pro). The exact model versions and deployment dates were recorded as part of the experimental configuration. All models were evaluated with temperature 0, schema-constrained output, identical prompt templates, and three repetitions per record. The primary analysis uses the held-out halves: 247 records comprising 190 violated and 57 clean cases. Because no prompt or metric definition was tuned after inspecting development results, the paper also reports all 494 records while explicitly treating the length baseline on the development halves as in-sample.

# V. Results

## A. Issue Detection

Table III summarizes the issue detection performance of the three LLM judges and two trivial baselines. On 247 held-out records, all three judges achieved high sensitivity (0.99–1.00), detecting nearly every injected defect in the controlled benchmark. However, this result should not be interpreted as evidence of equivalent sensitivity on real-world Oracle artifacts, given the synthetic dataset and its known surface-level correlations. However, specificity varied substantially (0.61–0.79), indicating differences in false-alarm behavior. The specification-length baseline achieved a balanced accuracy of 0.88, comparable to the LLM judges (0.81–0.89). Fig. 4 illustrates these results with 95% confidence intervals.

TABLE III. Issue detection by LLM judges and trivial baselines

| Evaluator | N | Defects detected | Clean passed | Sensitivity | Specificity | Balanced acc. |
|---|---|---|---|---|---|---|
| Held-out: GPT-4o | 247 | 188/190 | 43/57 | 0.99 [0.96,1.00] | 0.75 [0.63,0.85] | 0.87 [0.81,0.93] |
| Held-out: DeepSeek V4 Pro | 247 | 189/190 | 45/57 | 0.99 [0.97,1.00] | 0.79 [0.67,0.88] | 0.89 [0.84,0.94] |
| Held-out: Claude Sonnet 4.5 | 247 | 190/190 | 35/57 | 1.00 [0.98,1.00] | 0.61 [0.48,0.73] | 0.81 [0.75,0.87] |
| Held-out: Specification length | 247 | 182/190 | 46/57 | 0.96 [0.92,0.98] | 0.81 [0.69,0.89] | 0.88 [0.83,0.93] |
| Held-out: Always-issue | 247 | 190/190 | 0/57 | 1.00 | 0.00 | 0.50 |
| All: GPT-4o | 494 | 377/380 | 92/114 | 0.99 [0.98,1.00] | 0.81 [0.73,0.87] | 0.90 [0.86,0.93] |
| All: DeepSeek V4 Pro | 494 | 379/380 | 93/114 | 1.00 [0.99,1.00] | 0.82 [0.73,0.88] | 0.91 [0.87,0.94] |
| All: Claude Sonnet 4.5 | 494 | 380/380 | 77/114 | 1.00 [0.99,1.00] | 0.68 [0.58,0.75] | 0.84 [0.79,0.88] |
| All: Specification length | 494 | 363/380 | 101/114 | 0.96 [0.93,0.97] | 0.89 [0.81,0.93] | 0.92 [0.89,0.95] |
| All: Always-issue | 494 | 380/380 | 0/114 | 1.00 | 0.00 | 0.50 |

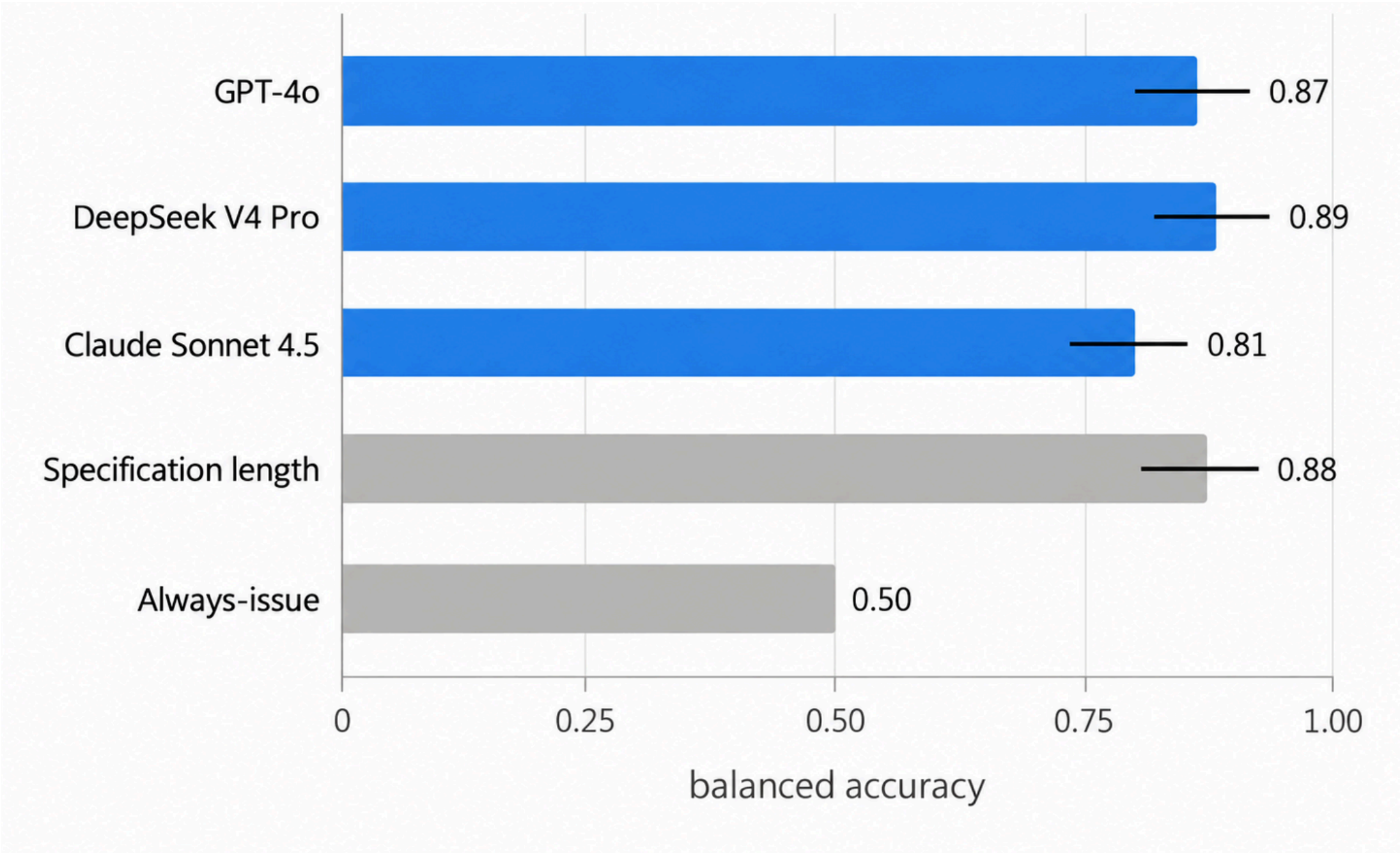

Fig. 4. Balanced accuracy of issue detection on the held-out halves (247 records). Error bars are 95% intervals. Blue bars are LLM judges; gray bars are trivial baselines.

Fig. 4 confirms that the observed differences in balanced accuracy are small relative to the reported uncertainty intervals. The following paired analysis examines whether these differences are statistically distinguishable. However, the specification-length baseline achieved a comparable score of 0.88, indicating that high detection accuracy alone does not establish source-grounded evaluation capability. The overlapping confidence intervals and paired statistical tests further show that none of the LLM judges significantly outperformed this simple baseline on the held-out dataset.

## B. *Paired Comparison with the Length Baseline*

The central empirical result is not that one LLM judge wins. The central empirical finding is that the benchmark does not provide sufficient statistical evidence to distinguish the LLM judges from a simple specification-length heuristic in binary issue detection. Although their error patterns differ, the observed balanced-accuracy differences are not statistically significant in the reported paired comparisons. On held-out records the length baseline reaches balanced accuracy 0.88 compared with 0.87, 0.89, and 0.81 for GPT-4o, DeepSeek V4 Pro, and Claude Sonnet 4.5. Exact paired tests in Table IV show no significant judge-versus-length difference on either record set.

Table IV reports the exact McNemar test results for paired classification correctness on the held-out and complete datasets.

TABLE IV. EXACT MCNEMAR TESTS: CORRECT BY FIRST ONLY / SECOND ONLY

| Comparison | Held-out (n=247) | All (n=494) |
|---|---|---|
| GPT-4o vs DeepSeek V4 Pro | 7 / 10, p=0.63 | 16 / 19, p=0.74 |
| GPT-4o vs Claude Sonnet 4.5 | 14 / 8, p=0.29 | 25 / 13, p=0.073 |
| DeepSeek V4 Pro vs Claude Sonnet 4.5 | 14 / 5, p=0.064 | 25 / 10, p=0.017 |
| GPT-4o vs specification length | 18 / 15, p=0.73 | 29 / 24, p=0.58 |
| DeepSeek V4 Pro vs specification length | 19 / 13, p=0.38 | 30 / 22, p=0.33 |
| Claude Sonnet 4.5 vs specification length | 17 / 20, p=0.74 | 28 / 35, p=0.45 |

The results indicate that none of the LLM judges significantly outperformed the specification-length baseline on the held-out dataset ($p > 0.05$). On the complete dataset, DeepSeek V4 Pro demonstrated higher classification accuracy than Claude Sonnet 4.5 in the unadjusted exact McNemar test ($p = 0.017$). However, this result does not remain statistically significant after Bonferroni correction for six pairwise comparisons (adjusted significance threshold of 0.0083). Consequently, the observed difference should be interpreted as exploratory rather than conclusive. These findings highlight the limitations of binary defect detection as an indicator of specification evaluation quality and motivate further analysis of category identification, affected-item localization, and evidence agreement.

## C. *Metric-Level False Alarms*

The shortcut arises from benchmark construction. In several completeness, coverage, and migration-risk metrics, the violated version was created by removing information, making it shorter than the corresponding clean specification. In additive or substitutive defects, such as unsupported claims or entity inconsistency, length is less predictive. Fig. 5 makes this visible through clean-pass counts per metric.

| Quality metric | GPT-4o | DeepSeek V4 Pro | Claude Sonnet 4.5 | Specification length |
|---|---|---|---|---|
| Ambiguous Statements | 5 | 6 | 2 | 6 |
| Behavioral Contradictions | 6 | 6 | 5 | 4 |
| Contradicted Claims | 6 | 6 | 5 | 6 |
| Control-Flow Completeness | 5 | 6 | 5 | 6 |
| Critical Source Fact Coverage | 3 | 4 | 4 | 6 |
| Data Transformation Completeness | 6 | 3 | 3 | 6 |
| Entity Consistency | 6 | 6 | 3 | 4 |
| Evidence Attachment | 6 | 6 | 4 | 4 |
| Hallucinated Entities | 6 | 6 | 6 | 3 |
| Interface Completeness | 3 | 4 | 4 | 6 |
| Risk Explanation Completeness | 5 | 2 | 4 | 6 |
| Semantic Migration Risks | 6 | 3 | 6 | 6 |
| Side-Effect Completeness | 5 | 4 | 6 | 6 |
| Source Dialect Leakage | 4 | 6 | 6 | 5 |
| Source Fact Coverage | 6 | 4 | 2 | 6 |
| Source Question Answerability | 2 | 3 | 4 | 6 |
| Terminology Drift | 5 | 6 | 2 | 6 |
| Unresolved References | 1 | 6 | 0 | 5 |
| Unsupported Claims | 6 | 6 | 6 | 4 |
| Total (out of 114) | 92 | 93 | 77 | 101 |
| Specificity (%) | 80.70 | 81.58 | 67.54 | 88.60 |

Fig. 5. Clean specifications correctly accepted by quality metric.

Fig. 5 reveals substantial metric-specific differences in false-alarm behavior. GPT-4o performs weakest on Unresolved References and Source Question Answerability, while DeepSeek V4 Pro exhibits its lowest clean-pass counts on Risk Explanation Completeness, Data Transformation Completeness, Semantic Migration Risks, and Source Question Answerability, indicating that false alarms are not confined to a single quality dimension. Claude Sonnet 4.5 exhibits particularly high false-alarm rates on Unresolved References, Ambiguous Statements, Source Fact Coverage, and Terminology Drift. In contrast, the specification-length baseline performs strongly on omission-oriented metrics because shorter specifications often correspond to injected defects. These patterns support metric-specific calibration rather than reliance on a single aggregate detection score.

### D. *Beyond Binary Detection*

The length baseline can exploit a surface cue but cannot identify a defect. Table V therefore reports structured-finding quality on violated records. These measures provide complementary evidence of the judges' ability to produce structured and localized findings beyond binary classification. However, category agreement and evidence overlap do not independently establish that the judgments rely on semantic analysis of the source code, particularly given the benchmark's synthetic construction and known shortcut signals.

TABLE V. Structured Finding Quality on Violated Records

| Measure | GPT-4o | DeepSeek V4 Pro | Claude Sonnet 4.5 |
|---|---|---|---|
| Held-out: labelled category among findings | 159/190 | 168/190 | 174/190 |
| Held-out: labelled category first | 147/190 | 147/190 | 145/190 |
| Held-out: affected item identified | 98/140 | 100/140 | 106/140 |
| Held-out: evidence overlap | 130/140 | 130/140 | 129/140 |
| Held-out: mean findings / violated | 1.28 | 1.49 | 1.77 |
| Held-out: mean findings / clean | 0.32 | 0.29 | 0.57 |
| All: labelled category among findings | 327/380 | 343/380 | 352/380 |
| All: labelled category first | 304/380 | 299/380 | 298/380 |
| All: affected item identified | 200/280 | 210/280 | 218/280 |
| All: evidence overlap | 251/280 | 265/280 | 256/280 |
| All: mean findings / violated | 1.28 | 1.54 | 1.77 |
| All: mean findings / clean | 0.25 | 0.26 | 0.43 |

On held-out violated records, the labelled category appears somewhere in 84%–92% of findings, the labelled affected item is identified in 70%–76% of applicable cases, and quoted evidence overlaps the labelled span in 92%–93%. These are qualitatively different from length-based detection because the evaluator must point to a category, object, or text span. Claude Sonnet 4.5 is the most exhaustive judge—it produces the most findings and most often includes the labelled category somewhere—but that exhaustiveness coincides with the highest false-alarm rate.

The benchmark therefore exposes a precision–thoroughness trade-off that balanced accuracy alone hides.

Terminology Drift requires one caveat: its allowed categories are named after the drifting terms, so the category vocabulary itself reveals candidate answers. Category-agreement results for that metric are consequently optimistic and should not be interpreted as independent evidence of diagnosis.

### E. Run Stability

Temperature 0 does not make the systems deterministic. On held-out records, the three runs disagree on 6 of 247 records for GPT-4o, 3 for DeepSeek V4 Pro, and 2 for Claude Sonnet 4.5. Over all 494 records the corresponding counts are 10, 7, and 5. Repeated runs are therefore warranted even in a nominally deterministic configuration, although majority voting changes only a small fraction of verdicts in this benchmark.

## VI. Discussion

### A. What the Experiment Does and Does Not Demonstrate

The findings expose an important distinction between high benchmark performance and validated evaluator capability. When a simple surface-level heuristic achieves comparable detection accuracy, the benchmark cannot independently establish whether an LLM judge relies on source semantics. Consequently, future specification assurance studies should prioritize shortcut-resistant datasets, localized evidence quality, and independent validation of evaluator findings.

This distinction is important for specification-driven migration. A production gate that simply predicts "something may be wrong" is useful only if it is well calibrated on clean artifacts. A review assistant that returns a typed finding with grounded evidence can support correction even when its binary sensitivity is saturated. The evaluation target should therefore move from verdict accuracy toward the quality of actionable findings.

### B. Updated Scientific Trajectory

The results support an updated trajectory for specification-driven migration research. The first stage establishes the specification as an intermediate representation for controlled Code–Text–Code evolution [1]. The second optimizes how that representation is carried through LLM contexts and across heterogeneous agents [2], [3]. The third extends explicit specification to deployment and security intent [4], [5], grounded in lifecycle risk models [6], [7]. The present work adds a fourth stage: specification assurance must itself be benchmarked under leakage-resistant, shortcut-aware conditions before specifications can be treated as trustworthy generation contracts.

This trajectory suggests that future migration pipelines should separate four responsibilities: extraction of source facts, generation of a technology-neutral specification, evidence-linked evaluation of that specification, and only then target-dialect synthesis. The evaluator should be replaceable: deterministic fact extraction, question answering, embeddings, or LLM judges can be compared through the same benchmark interface. This is scientifically preferable to coupling the benchmark to a single judge architecture.

### C. Implications for Benchmark Design

Three benchmark-design requirements follow directly. First, clean and defective records should be matched for length and other obvious surface statistics, otherwise a trivial classifier can exploit the construction procedure. Second, ground truth must be excluded from every inference channel, including identifiers, filenames, logs, metadata, and examples. Third, binary labels are insufficient when the benchmark contains only one injected defect: additional true findings cannot be distinguished from false positives. Future datasets should use expert-validated multi-defect annotations and independent evidence spans.

### D. Implications for Migration Engineering

For migration engineering, a useful evaluator should minimize two asymmetric risks. False negatives allow corrupted specifications to drive target generation; false positives increase review cost and can block otherwise sound artifacts. In the present benchmark sensitivity is already 0.99–1.00, so the practical optimization target becomes specificity and finding precision. This is consistent with the broader LLM-as-a-judge literature, where permissive judges tend to appear thorough while over-reporting [14], [15], [34].

Evidence attachment is particularly important. A finding that quotes the source span or specification phrase it relies on can be inspected, corrected, and linked to a migration task. This supports a traceable workflow in which the specification is not merely scored but repaired before it becomes input to target-dialect generation. Such a workflow is aligned with the representation-driven migration model [1]–[3] and with security knowledge transition, where requirements are expected to remain traceable through implementation and verification [5].

## VII. Limitations

The benchmark is a silver-standard synthetic dataset. Sources, specifications, and labels have not been exhaustively verified by independent database experts; a clean record may contain an unlabelled defect and an injected label may occasionally be debatable. Some false alarms could therefore be genuine findings. The artifacts are also small, self-contained Oracle examples, whereas industrial objects depend on schemas, packages, session state, external routines, privileges, and cross-file data flow.

There are only 13 held-out records per metric, including three clean records and one held-out record per issue category. Metric-level results are therefore descriptive. Pooled inference assumes the 19 metrics are sufficiently comparable. Furthermore, repeated use of source artifacts across metric datasets may introduce dependencies between observations. Although source-sharing records were assigned to the same data partition, the pooled confidence intervals and paired tests do not explicitly model clustering by source artifact. Consequently, the reported uncertainty estimates should be interpreted with this limitation in mind. Specificity has wider uncertainty than sensitivity because the benchmark contains only 57 held-out clean records.

The length confound is a known property of the data and is deliberately retained in the reported experiment because discovering it is part of the methodological result. It prevents interpreting binary detection as evidence of semantic reasoning. A revised benchmark should construct length-matched adversarial clean/defective pairs and include negative controls for other superficial signals.

Only one family of evaluator—LLM judges that read source and specification—was compared. Deterministic fact extraction, fact-grounded hybrids, bidirectional question answering, and embedding-based methods remain future comparisons. Finally, the metric definitions were used as first written; at least Source Question Answerability and Unresolved References appear mismatched to some clean cases, and development-data calibration may reduce false alarms.

## VIII. Conclusion

LRFB provides a leakage-resistant framework for measuring evaluators of AI-generated database specifications. Across 19 metrics and 494 labelled Oracle PL/SQL source–specification pairs, three LLM judges detect nearly every injected defect. On the 247 held-out records their balanced accuracy ranges from 0.81 to 0.89, but a threshold on specification length reaches 0.88 and is statistically indistinguishable from every judge. Binary detection therefore cannot establish source-grounded evaluation on this benchmark.

The stronger result appears beyond the verdict. On held-out violated records, the judges include the labelled defect category in 84%–92% of cases and quote evidence overlapping the labelled span in 92%–93%. These outputs can support review and correction in a way a shortcut baseline cannot. The next stage should therefore rebuild the benchmark with expert validation, length-matched records, more clean cases, and multi-defect annotations, then compare LLM judges with deterministic, question-answering, hybrid, and embedding evaluators through the same sealed protocol. In a specification-driven migration pipeline, the intermediate specification should be treated as a generation contract only after its evaluator has survived that level of validation.